\documentclass[aps,prd,twocolumn,showpacs,nofootinbib]{revtex4-1}
\usepackage[latin1]{inputenc}
\usepackage{latexsym}
\usepackage{amsmath,amsfonts}
\usepackage{amsbsy}
\usepackage{mathrsfs}
\usepackage{color}
\usepackage{psfrag}
\usepackage{enumerate}
\usepackage{amsmath,amssymb,calc,amsfonts}
\usepackage{latexsym}
\usepackage[colorlinks=true,linkcolor=blue,urlcolor=blue,citecolor=red]{hyperref}
\usepackage{amsmath}
\usepackage{amsfonts}
\usepackage{amssymb}

\usepackage{latexsym}
\usepackage{amsfonts}
\usepackage{t1enc}
\usepackage{times}
\usepackage{xmpmulti}
\usepackage{color}
\usepackage{bm}
\usepackage{hyperref}

\font\tenscr=rsfs10 scaled1100
\font\sevenscr=rsfs7 
\font\fivescr=rsfs5 
\skewchar\tenscr='177
\skewchar\sevenscr='177
\skewchar\fivescr='177
\newfam\scrfam
\textfont\scrfam=\tenscr
\scriptfont\scrfam=\sevenscr
\scriptscriptfont\scrfam=\fivescr

\def\scri{{\fam\scrfam I}}
\def\scrm{{\fam\scrfam M}}

\def\scrl{{\fam\scrfam L}}
\def\scrt{{\fam\scrfam T}}

\begin{document}

\title{Center of Mass and spin for isolated sources of gravitational radiation}
\date{\today}

\author{Carlos N. Kozameh$^\dagger$, Gonzalo D. Quiroga$^\dagger$}

\affiliation{$^\dagger$ Instituto de F\'isica Enrique Gaviola, FaMAF, Universidad Nacional de C\'ordoba, C\'ordoba, Argentina.}

\begin{abstract}
We define the center of mass and spin of an isolated system in General Relativity. The resulting relationships between these variables and the total linear and angular momentum of the gravitational system are remarkably similar to their Newtonian counterparts, though only variables at the null boundary of an asymptotically flat spacetime are used for their definition.
We also derive equations of motion linking their time evolution to the emitted gravitational radiation. The results are then compared to other approaches. In particular one obtains unexpected similarities as well as some differences with results obtained in the Post Newtonian literature .

These equations of motion should be useful when describing the radiation emitted by compact sources such as coalescing binaries capable of producing gravitational kicks, supernovas, or scattering of compact objects.
\end{abstract}

\maketitle

\section{Introduction}

\quad The main goal of this work is to define the notions of center of mass and intrinsic angular momentum for isolated systems and obtain their dynamical evolution when gravitational radiation is emitted. The evolution of isolated systems and its  gravitational radiation is naturally described using the notion of asymptotically flat spacetimes. Thus, our approach will be based on this mathematical framework.

Both in Newtonian theory and special relativity one can find a particular trajectory with the property that the mass dipole moment vanishes at this trajectory. This special trajectory is called the center of mass. If one would like to generalize this concept to GR, then the goal would be to find a worldline in spacetime with analogous properties to the one described in Newtonian gravity or special relativity. The first step is therefore to provide an adequate definition of mass dipole moment in GR.

One also expects that any suitable definition of center of mass should be related to other global quantities like the Bondi mass $M$ or momentum $P^i$ by the relation $P^i=M V^i+$ \emph{radiation terms}. However, in contrast to Newton's theory of gravity, the Bondi mass or momentum will not be conserved for an isolated system since gravitational waves carry away mass and momentum. Therefore, one also expects that the velocity of the center of mass will change when radiation is emitted.

It is also worth mentioning that there is a qualitative difference between the geometrical meaning of the dipole mass moment in Newtonian gravity and in special relativity. Whereas in Newton theory the mass moment is a vector, in special relativity it is a component of the so called, the mass dipole moment/angular momentum 2-form \cite{Pen2}. Thus, to implement this program one should generalize the mass dipole moment/angular momentum 2 form to GR, and then define the center of mass worldline as the special place where the mass dipole vanishes. As a bonus one should obtain the intrinsic angular momentum evaluating the non-vanishing part of this generalized 2-form on the center of mass worldline.

However, as one can see in the literature, there are many definitions of angular momentum/mass dipole moment for isolated systems in general relativity. As a non complete list of authors we could mention Dray and Streubel \cite{DrayStreubel}, Bramson \cite{Bramson}, Geroch \cite{Geroch}, Helfer \cite{Helfer}, Moreschi \cite{Moreschi}, Penrose \cite{Penrose} and Winicour \cite{Winicour}. Although a recent living review \cite{Szab} offers a complete survey of the main results in the field  with the main motivations and technical aspects of each definition, the fact that there is no agreement among these alternative approaches reflects the difficulty of the subject. However, there is a common link between them that can be used as a starting point: all the approaches agree for quadrupole radiation.

This fact has been used in the Adamo-Newman-Kozameh \cite{ANK} approach. By restricting themselves to quadrupole radiation data, it is shown that both the center of mass and angular momentum are defined from an asymptotic Weyl scalar whose $l=1$ part of the spherical harmonic decomposition transforms as a 4-dim two form under the action of the homogeneous Lorentz algebra of the BMS group, the available kinematic geometry of null infinity \cite{ANK}. Moreover, the ANK formulation is the only one that gives equations of motion for both the center of mass and spin of an isolated system.
In the ANK approach the center of mass and spin are respectively the real and imaginary parts of a complex worldline defined in the solution space of the good cut equation. The geometrical interpretation of this space is that each solution describes a congruence of asymptotically shear free null geodesics reaching null infinity. The novelty of the formulation lies in the definition of the spin as an intrinsic property of this complex worldline and thus it can be used to give a classic definition of a gravitational particle with spin.

 From our perspective, however, the ANK approach has some points that deserve further attention
\begin{enumerate}
 \item the angular momentum is only defined for quadrupole radiation and cannot be extended  to generic radiation since it does not give the expected results when the space time has a rotational symmetry.   Thus, one must generalize this definition to spacetimes with arbitrary gravitational radiation, and include the case when the spacetime is axially symmetric,
 \item By assumption, the approach is based on null congruences with vanishing shear at null infinity. However, at null infinity the shear of the future null cone of any point does not vanish. This follows from the optical equations since a non vanishing Weyl curvature on the null cone induces non vanishing shear. Thus, the center of mass worldline defined on the solution space of asymptotically vanishing shears does not correspond to a worldline of the underlying spacetime.
 \item The spin is defined as the imaginary part of a complex worldline instead of simply evaluating the angular momentum at the center of mass.
\end{enumerate}

In this work we present new definitions of center of mass and spin using the available tools on asymptotically flat spacetimes. In these new definitions we try to answer the above issues by constructing one parameter (Newman-Unti) foliations of null infinity that are related to null cones cuts from points of the spacetime and have non vanishing shear at null infinity. The spin is simply the angular momentum at the center of mass and the center of mass is the place where the mass dipole moment vanishes.

Since there are many technical details, some of them involved, it is better to outline here the main ideas of our approach. In this way the reader can have a broad picture without the technical complications.

We first introduce the notion of null cone cuts as the intersection of the future null cones from points $x^a$ of the spacetime with null infinity. We then define the regularized null cone cuts (or RNC cuts) as the Huygens part of the null cone cuts. By construction the RNC cuts are smooth 2-surfaces at null infinity that parametrically depend on the points of the spacetime. If the points $x^a(u)$ describe a worldline the RNC cuts yield a special Newman-Unti (NU) family of cuts.

We then introduce the notion of linkages \cite{Winicour,GW} on this special family of NU cuts to define the dipole mass moment/angular momentum. The main reason for this choice is that the linkage is a linear generalization of the Komar formula which automatically yields the standard Komar definition when the spacetime has a Killing field associated to a rotational symmetry. By restricting the linkages to the RNC cuts we fix one of the main problems in the linkage formulation. Instead of having a definition of angular momentum with a supertranslation freedom we restrict the freedom to the RNC family, a special 4-dim family of Newman-Unti cuts, where the notions of dipole mass moment and angular momentum are introduced (see ref. \cite{Szab} page 30). Although there are still infinite degrees of freedom, one for each worldline, the freedom is analogous to the choice of origins in the Newtonian definition of angular momentum.

Finally, by demanding that on one RNC cut the mass dipole term vanishes we select a special point associated with this cut that by definition is called center of mass. Evaluating the angular momentum on this special RNC cut yields the intrinsic angular momentum or spin.

Note that the notion of a null cone cut as the intersection of null cones from points of the spacetime with null infinity is purely geometric. Note also that, as pointed out by Geroch and Winicour, the linkages also offer a coordinate free definition. Thus in principle our construction solely depends on a family of NU cuts and it is independent on the coordinates used for its description.

As it was also done in the ANK formulation, this approach yields explicit equations of motion for the center of mass and spin when gravitational radiation is emitted from the source. The equations of motion of both formulations can be compared and, as one would expect, they are different. Given that there is available in the literature models of binary coalescence based on the post Newtonian approximation it is also of great interest to compare our equations with these models. It is surprising to find out that the time evolution of the total mass, linear and angular momentum in our approach agrees with the PN formulation up to octupole terms in the gravitational radiation.

It is left for future work to analyse other definitions of dipole mass moment/angular momentum, that yield the Komar formula for axial symmetry.  In this sense one should mention that the Gallo-Moreschi definition \cite{GM}, following a completely different approach, gives exactly the same formula as the Linkages on Bondi sections. (The old definition had some freedom and the original way to fix it yielded a different result \cite{Moreschi2}.)

Since the Moreschi approach also defines a preferred family of Bondi cuts (called nice sections) it is worth making a few remarks about them. The nice sections are found by demanding that the $l\geq2$ part of the supermomentum at null infinity vanishes when restricted to those cuts. The nice section equation is obtained and the center of mass frame is a special solution of the equation. The nice section equation is different from either the null cone cut equation at a local level or the regularized null cone cut equation at a global level on the sphere. Whereas the cut equations  have (at least) a linear dependence on the Bondi shear, the nice section equation has a quadratic dependence. In addition, since the RNC cut equation yield monoparametric families of NU cuts whose areas are time dependent and in general are not unit spheres, the nice sections are by construction Bondi surfaces and thus have unit area.
Furthermore, the solutions to the nice sections are specially adapted to get rid of unwanted supermomentum terms and thus define unambiguously the notion of center of mass and intrinsic angular momentum at each Bondi time. On the other hand, our formulation is based on a special monoparametric family of cuts at null infinity together with a coordinate free approach to find the notion of center of mass and intrinsic angular momentum. In the end however, one uses a Bondi coordinate system to obtain explicit description of the approach. Therefore, it is worthwhile to examine in more detail both approaches and find similarities and differences in a future work.

The technical material needed for this work is presented in Sections 2-4. Section 5 is the main part of this work. We give definitions of center of mass and spin, derive the equations of motion and compare our results with other approaches. The work ends with some concluding remarks.

\section{Foundations}
In this section, we introduce several of the key ideas and the basic tools that are needed for our later discussion.

\subsection{Asymptotically flatness and $\scri^+$}
We first introduce some mathematical framework. In particular we introduce the notion of an isolated source of gravitational radiation realized by defining the so called asymptotically flat spacetimes. Bondi, Sachs and collaborators in the sixties \cite{BBM,sachs3}, used a canonical coordinate system were mass, momentum and gravitational radiation could be defined. Later, Penrose gave a geometrical definition using a rescaled metric together with a null boundary \cite{Pen}. Both approaches can be found in the review of Newman and Tod \cite{ntod}. We follow Newman and Tod in the following definitions.

A spacetime $(\scrm, g_{ab})$ is called asymptotically flat if the curvature tensor vanishes as infinity is approached along the future-directed null geodesics of the spacetime. These geodesics end up at what is referred to as future null infinity $\scri ^{+}$, the future null boundary of the spacetime . These ideas can be formalized by giving the following,

\emph{Definition:} a future null asymptote is a manifold $\hat \scrm$ with boundary $\scri^+ \equiv \partial \hat\scrm$ together with a smooth lorentzian metric $\hat{g}_{ab}$, and a smooth function $\Omega$ on $\hat\scrm$ satisfying the following
\begin{itemize}
  \item $\hat{\scrm}=\scrm \cup \scri^+$
  \item On $\scrm$, $\hat{g}_{ab}=\Omega ^2 g_{ab}$ with $\Omega >0$
  \item At $\scri^+$, $\Omega=0$, $n_a^{\ast} \equiv \partial _a \Omega \neq 0$ and $\hat {g}^{ab}n_a^{\ast}n_b^{\ast}=0$
\end{itemize}
We assume $\scri^{+}$ to have topology $S^2\times R$. A Newman-Unti (N-U) coordinate system \cite{nu} is introduced in the neighborhood of $\scri ^{+}$,as follows. We first give a regular one-parameter family of closed 2-dim cuts at null infinity, labelled by the parameter $u$ which meet every generator once. The stereographic coordinates $(\zeta,\bar \zeta)$ label each generator on the cut. We then construct a family of null surfaces whose intersection with $\scri$ are these NU cuts,and use the affine parameter  $r$ on each null surface as our last coordinate.

Since $\scri^+$ is a null hypersurface in the rescaled manifold $\hat{\scrm}$ the restriction of the rescaled metric on this null boundary takes the form
\begin{equation}
d\hat{s}^2=\frac{4d\zeta d\bar \zeta}{P^2}.
\end{equation}
with $P(u,\zeta,\bar \zeta)$ a strictly positive function.  With the choice of $\Omega=r^{-1}$ as the conformal factor, the physical metric is then given as
\begin{equation}\label{metrics2}
ds^2=\frac{4 r^2 d\zeta d\bar \zeta}{P^2}.
\end{equation}

\subsection{Null Tetrads and Operators on the sphere}
Associated with the NU coordinates $(u,r,\zeta,\bar\zeta)$, there is a null tetrad system denoted by ($l_{a}^{\ast}$,$n_{a}^{\ast}$,$m_{a}^{\ast}$,$\bar{m}_{a}^{\ast}$). The first null tetrad covector $l_{a}^{\ast}$ is defined as \cite{ntod}
\begin{equation} \label{ladef}
l_{a}^{\ast}=\nabla _{a}u,
\end{equation}
Thus, $l^{a\ast}$ is a null vector tangent to the  null surface $u= const.$. The remaining null vectors are then prescribed at $\scri^+$ and then parallel propagated inwards along $l^{a\ast}$. The second tetrad vector $n^{\ast a}$ is tangent to the null generators of $\scri^+$  and normalized to $l^{\ast a}$
\begin{equation} \label{ln}
n^{a\ast}l^\ast_a=1.
\end{equation}
The null tetrad at $\scri^+$ is finally completed by selecting two complex null vectors at the intersection of $u=const.$ and $\Omega=0$. The complex vector $m^{a\ast}$ orthogonal to $l^{a\ast}$ and $n^{a\ast}$ is normalized to
\begin{equation}
m_{a}^{\ast} \bar{m}^{\ast a}=-1.
\end{equation}

The null tetrad for the spacetime is then constructed from parallel propagation along $l^{a\ast}$. The spacetime metric is given by
\begin{equation} \label{metrica}
g_{ab}=l_{a}^{\ast}n_{b}^{\ast}+n_{a}^{\ast}l_{b}^{\ast}-m_{a}^{\ast}\bar{m}_{b}^{\ast}-\bar{m}_{a}^{\ast}m_{b}^{\ast}.
\end{equation}
(In this work the letters $a,b,c,d$ take values $0,1,2,3$.) For more details on the asymptotic form of the metric in NU coordinates see ref. \cite{ntod}.

Since there is a gauge freedom in the choice of conformal factor $\Omega$ one can freely choose the function $P(u,\zeta,\bar \zeta)$. The particular choice $P=P_0=(1+\zeta \bar{\zeta})$, yields a two-surfaces metric (\ref{metrics2}) of unit radius that is Lie derived along the null directions of $\scri^+$. For this particular choice of conformal factor a Bondi time $u_B$ is introduced as the affine length of the null geodesic $n^a \equiv {\hat g}^{ab}\nabla_b \Omega$. The covector $l_{a}=\nabla _{a}u_B$ yields a Bondi tetrad $(l_a,n_a,m_a,\bar{m}_a)$ following the same procedure as above.

Since $u_B= const.$ are unit spheres whereas $u=const.$ are not, the description of one cut in terms of the other may be written as
\begin{eqnarray}
u_{B}&=&Z(u, \zeta , \bar{\zeta}),\label{ubz}\\
u&=&T(u_B,\zeta,\bar{\zeta}). \label{ut}
\end{eqnarray}
where $Z$ is a smooth function and $T$ is the inverse of $Z$. They satisfy $\dot{T}Z^{\prime}= 1$, where "dot" and "prime" denote the derivative with respect to $u_B$ and $u$ respectively.

We also introduce the concept of spin weight. A quantity $\eta$ that transforms as $\eta \rightarrow e^{is\lambda }\eta$ under a rotation $m^{a\ast} \rightarrow e^{i\lambda }m^{a\ast }$ is said to have a spin weight $s$. For any function $f(u, \zeta, \bar{\zeta})$, we define the differential operators $\eth^{\ast}$ and $\bar{\eth }^{\ast}$ \cite{ANK} by
\begin{eqnarray}
\eth^{\ast} f&=&P^{1-s}\frac{\partial (P^{s}f)}{\partial \zeta }, \label{eth}\\
\bar{\eth }^{\ast}f&=&P^{1+s}\frac{\partial (P^{-s}f)}{\partial \bar{\zeta}}, \label{ethb}
\end{eqnarray}
where $f$ has a spin weight $s$ and $P$ is the conformal factor defining the metric (\ref{metrics2}). Likewise, we define
\begin{eqnarray}
\eth f&=&P_0^{1-s}\frac{\partial (P_0^{s}f)}{\partial \zeta },\\
\bar{\eth }f&=&P_0^{1+s}\frac{\partial (P_0^{-s}f)}{\partial \bar{\zeta}}, \label{ethb}
\end{eqnarray}
with $P_0=(1+\zeta \bar{\zeta})$. Furthermore, using $P=P_0 Z'$ (which follows from $\frac{r_B}{P_0}= \frac{r}{P}$ and $r_B=Z^{\prime}r$ \cite{nu}) one can relate these two operators as
\begin{eqnarray}
\eth ^{\ast }f&=&Z^{\prime }\eth f+sf\eth Z^{\prime } \\
\bar{\eth }^{\ast }f&=&Z^{\prime }\bar{\eth }f-sf\bar{\eth }Z^{\prime }.
\end{eqnarray}
The above equation (which is not a coordinate transformation between the NU and Bondi coordinate systems) will be used below to expand regular functions on the sphere in the standard spherical harmonic basis.

Now, we are interested in the relationship between the NU and Bondi null tetrads. We start by rewriting eq. (\ref{ladef}) in the form $l_{a}=\nabla _{a}Z(u,\zeta ,\bar{\zeta})$ and using the orthogonality of the null vectors to get
\begin{eqnarray}
l_{a}^{\ast } &=&\frac{1}{Z^{\prime }}[l_{a}-\frac{L}{r_{B}}\bar{m}_{a}-%
\frac{\bar{L}}{r_{B}}m_{a}+\frac{L\bar{L}}{r_{B}^{2}}n_{a}], \label{la}\\
n_{a}^{\ast } &=&Z^{\prime }n_{a}, \label{na}\\
m_{a}^{\ast } &=&m_{a}-\frac{L}{r_{B}}n_{a} \label{ma}, \\
\bar{m}_{a}^{\ast } &=&\bar{m}_{a}-\frac{\bar{L}}{r_{B}}n_{a}, \label{mba}
\end{eqnarray}
where
$$L(u_{B},\zeta ,\bar{\zeta })=\eth Z(u,\zeta ,\bar{\zeta }). $$

\subsection{The Spin Coefficient Formalism}
In this subsection we will describe the NP formalism in term of the Bondi coordinates $(u_B,r_B,\zeta,\bar\zeta)$, this means that all introduced functions depend on these coordinates.
First, we introduce the Ricci rotation coefficients $\gamma_{\mu \nu \rho}$ \cite{ntod,np}
\begin{equation}
\gamma_{\mu \nu \rho}={\lambda ^{a}}_{\rho}{\lambda ^{b}}_{\nu}\nabla _{a}\lambda_{b \mu},
\end{equation}
the Ricci rotations coefficients satisfy
\begin{equation}
\gamma _{\mu \nu \rho}=-\gamma _{\nu \mu \rho}.
\end{equation}%
where
\begin{equation}
{\lambda^a}_{\mu}=(l^a,n^a,m^a,\bar{m}^a),
\end{equation}
where $\mu,\nu,\rho=1,2,3,4$ are tetrad indexes. The twelve spin coefficients are defined as combinations of the $\gamma_{\mu \nu \rho}$
\begin{eqnarray}
\alpha  &=&\frac{1}{2}(\gamma _{124}-\gamma _{344});\quad \lambda =-\gamma
_{244};\quad \kappa =\gamma _{131}  \nonumber  \label{spincoef} \\
\beta  &=&\frac{1}{2}(\gamma _{123}-\gamma _{343});\quad \mu =-\gamma
_{243};\quad \rho =\gamma _{134} \\
\gamma  &=&\frac{1}{2}(\gamma _{122}-\gamma _{342});\quad \nu =-\gamma
_{242};\quad \sigma =\gamma _{133}  \nonumber \\
\varepsilon  &=&\frac{1}{2}(\gamma _{121}-\gamma _{341});\quad \pi =-\gamma
_{241};\quad \tau =\gamma _{132}  \nonumber
\end{eqnarray}
The Peeling theorem of Sachs \cite{sachs} tell us the asymptotic behavior of the spin coefficients \cite{ANK}.
\begin{eqnarray}
\kappa  &=&\pi =\varepsilon =0;\qquad \rho =\bar{\rho};\qquad \tau =\bar{%
\alpha}+\beta   \nonumber  \label{spincoef-des-r} \\
\rho  &=&-r_B^{-1}-\sigma ^{0}\bar{\sigma}^{0}r_B^{-3}+O(r_B^{-5})  \nonumber \\
\sigma  &=&\sigma ^{0}r_B^{-2}+[(\sigma ^{0})^{2}\bar{\sigma}^{0}-\psi
_{0}^{0}/2]r_B^{-4}+O(r_B^{-5})  \nonumber \\
\alpha  &=&\alpha ^{0}r_B^{-1}+O(r_B^{-2})  \nonumber \\
\beta  &=&\beta ^{0}r_B^{-1}+O(r_B^{-2}) \\
\gamma  &=&\gamma ^{0}-\psi _{2}^{0}(2r_B^{2})^{-1}+O(r_B^{-3})  \nonumber \\
\mu  &=&\mu ^{0}r_B^{-1}+O(r_B^{-2})  \nonumber \\
\lambda  &=&\lambda ^{0}r_B^{-1}+O(r_B^{-2})  \nonumber \\
\nu  &=&\nu ^{0}+O(r_B^{-1})  \nonumber
\end{eqnarray}%
where the relationships among the r-independent functions
\begin{eqnarray}
\alpha ^{0} &=&-\bar{\beta}^{0}=-\frac{\zeta }{2},\qquad \gamma ^{0}=\nu
^{0}=0,   \nonumber \\
\omega ^{0} &=&-\bar{\eth }\sigma ^{0}, \qquad \lambda ^{0}=\dot{\bar{\sigma}}%
^{0},\qquad \mu ^{0}=-1,  \nonumber
\end{eqnarray}
with $\sigma^0$ the value of the Bondi shear at null infinity. This complex scalar is called the Bondi free data (or Bondi news) since $\ddot{\sigma}^0$ yields the gravitational radiation reaching null infinity. Since the BOndi shear is a s.w. 2 object it can be written as
$$\sigma^0 = \eth^2(\sigma_R + i \sigma_I).$$

The real functions $\sigma_R$, $\sigma_I$ are respectively called the electric and magnetic part of the Bondi shear. They are related to the mass and magnetic nth-poles moments of the gravitational source.

As the spacetime is assumed to be empty in a neighborhood of $\scri^+$ the gravitational field is given by the Weyl tensor. Using the available tetrad one defines five complex scalars, whose asymptotic behavior is \cite{sachs}
\begin{eqnarray*}
\psi_{0} &=&{C_{abc}}^dm^{a}l^{b}l^{c}m_{d}\simeq \frac{\psi _{0}^{0}}{r_B^{5}}, \quad \psi _{3} ={C_{abc}}^dl^{a}n^{b}n^{c}\bar{m}_{d}\simeq \frac{\psi _{3}^{0}}{r_B^{2}}.\\
\psi_{1} &=&{C_{abc}}^dn^{a}l^{b}l^{c}m_{d}\simeq \frac{\psi _{1}^{0}}{r_B^{4}}, \quad \psi _{4} ={C_{abc}}^d\bar{m}^{a}n^{b}n^{c}\bar{m}_{d}\simeq \frac{\psi_{4}^{0}}{r_B}.\\
&& \psi _{2} =\frac{1}{2}({C_{abc}}^dl^{a}n^{b}m^{c}\bar{m}_{d}-C_{abcd}l^{a}n^{b}l^{c}n_{d})\simeq\frac{\psi _{2}^{0}}{r_B^{3}}.
\end{eqnarray*}

Using the peeling theorem the radial part of the Einstein equations can be integrated leaving only the Bianchi identities at $\scri$ as the unsolved equations. In a Bondi frame the resulting equations look remarkably simple. Some of those equations relate the Weyl scalars with the Bondi shear, i.e., \cite{ANK,ntod}
\begin{eqnarray}
\psi_{2}^{0}+\eth^{2}\bar{\sigma }^{0}+\sigma ^{0}\dot{\bar{\sigma }}^{0}&=&\bar{\psi }_{2}^{0}+\bar{\eth }^{2}\sigma ^{0}+\bar{\sigma }^{0}\dot{\sigma }^{0},\label {psi2}\\
\psi_{3}^{0}&=&\eth \dot{\bar{\sigma}}^0,\\
\psi_4^{0}&=&-\ddot{\bar{\sigma}}^0,
\end{eqnarray}
Here the $\eth$ operator is taken at $u_B=const$.

In the same way we can define the Weyl scalars in N-U using the fact that the Weyl tensor ${C_{abc}}^d$ is conformally invariant \cite{ntod}.
\begin{eqnarray*}
\psi _{1}^{\ast} &=&{C_{abc}}^d n^{a\ast}l^{b\ast}l^{c\ast}m_{d}^{\ast}\simeq \psi _{1}^{0\ast}r^{-4},\\
\sigma^{\ast} &=&m^{\ast a}m^{\ast b}\nabla _{a}l_{b}^{\ast }\simeq\sigma ^{0\ast}r^{-2}.
\end{eqnarray*}
From the equations (\ref{la}-\ref{mba}) we can find transformations from NU to Bondi for any scalar or spin coefficient \cite{AN,KQ}. In particular we are interested in
\begin{equation}\label{transformacion}
\frac{{\psi }_{1}^{0\ast}}{Z^{\prime 3}}=[\psi _{1}^{0}-3L\psi _{2}^{0}+3L^{2}\psi_{3}^{0}-L^{3}\psi _{4}^{0}],
\end{equation}
where ${\psi }_{1}^{0\ast}$ is constructed from the N-U tetrad. Similarly we find the relation between $\sigma^{0\ast}$ and $\sigma^{0}$ \cite{AN}
\begin{equation}\label{sigma*}
\frac{\sigma^{0\ast }}{Z^{\prime}}=\sigma ^{0}-\eth^{2}Z.
\end{equation}
where $\sigma^{0\ast}$ is the NU shear \cite{nu}.

\subsection{Evolution equations}
Finally, the Bianchi identities (in Bondi coordinates) are given
by \cite{ANK,ntod}
\begin{eqnarray}
\dot{\psi}_{0}^{0}&=&-\eth \psi_1^0+3\sigma ^{0}\psi_2^0,\label{psi0prima}\\
\dot{\psi}_{1}^{0}&=&-\eth \psi_2^0+2\sigma ^{0}\psi_3^0,\label{psi1prima}\\
\dot{\psi}_2^0&=&-\eth \psi_3^0 +\sigma^0\psi_4^0. \label{Psi_2dot}
\end{eqnarray}
Note that eq. (\ref{psi2}) defines a real variable $\Psi$ called the mass aspect \cite{BBM}.
\begin{equation} \label{asp.masa}
\Psi =\psi _{2}^{0}+\eth ^{2}\bar{\sigma }^{0}+\sigma ^{0}\dot{\bar{\sigma }}^{0},
\end{equation}
In term of $\Psi$ is possible to write the Bondi Mass $M$ and Bondi lineal momentum $P^i$ by
\begin{eqnarray}
M &=&-\frac{c^{2}}{8\pi \sqrt{2}G}\int \Psi dS,\\
P^{i} &=&-\frac{c^{3}}{8\pi \sqrt{2}G}\int {\Psi }\tilde{l}^{i} dS,
\end{eqnarray}
with
\begin{equation}
\tilde{l}^{i}=\frac{1}{1+\zeta \bar{\zeta}}(\zeta +\bar{\zeta},-i(\zeta -%
\bar{\zeta}),1-\zeta \bar{\zeta}).
\end{equation}
with $dS=\frac{4d\zeta \wedge d\bar\zeta}{P_0^2}$ the area element on the unit sphere and where $i,j,k,l,m=1,2,3$ are three dimensional Euclidian indices. It is important to note that at $\scri$ we move upstairs and downstairs indices with the flat metric.

It is also quite convenient to give the evolution equation for $\Psi$. Directly from eq. (\ref{Psi_2dot}) one obtains
\begin{equation}
\dot{\Psi}=\dot{\sigma} ^{0}\dot{\bar{\sigma }}^{0}.\label{psiprima}
\end{equation}
This equation will be used later.
\section{Regularized Null Cone Cuts}
Another important construction in this work is a special NU foliation obtained from the null cone cuts of null infinity or NC cuts for short.

Given a point $x^a$ on the spacetime and denoting by $N_x$ the future null cone from $x^a$, we define a null cone cut (NC cut) as $N_x \cap \scri ^{+}$. The local and global properties of the NC cuts have been extensively analysed \cite{KLR,IKR,FNS} and some of them are summarized in the Appendix. In this section we briefly review some results that are needed for this work.

In flat spacetime the NC cuts are smooth surfaces that can be written as a regular functions on the sphere, i.e.,

\begin{equation} \label{flat cut}
Z_0= x^a\ell_a, \qquad x^a= (t,x^{i}), \qquad \ell_a=(Y_0^0,-\frac{1}{2}Y^0_{1i}),
\end{equation}
with $x^a$ the apex of the null cone and $Y_0^0,Y^0_{1i}$ the $\ell =0,1$ spherical harmonics. If the apex describes a timelike worldline $x^a(\tau)$ in Minkowski space the NC cuts describe a one parameter foliation of Null Infinity.

The idea is to generalize this concept for asymptotically flat spacetimes. This is a highly non trivial task since curvature induces caustics on the future null cones of points. Thus, the NC cuts have self-intersections and caustics. Nevertheless one can show that it is always possible to find a neighborhood at null infinity where a NC cut is a smooth 2-surface. In a Bondi coordinate system, this surface is a graph of a function

\begin{equation} \label{nc cut}
u_B=Z(x^a,\zeta, {\bar \zeta}).
\end{equation}

For the type of systems we are interested in describing, i.e., gravitational radiation coming from compact sources in the observation volume of aLIGO, one can always assume the null cone cut can be described by the above function. Moreover, to recover the point $x^a$ from which the radiation is coming one does not need the whole 2-surface, rather a small  neighborhood of points $(\zeta,{\bar \zeta})$ in the sphere. This follows from the dual meaning of $Z$ as the past null cone from $(u_B, \zeta, {\bar \zeta})$. Thus, $(\eth Z, {\bar \eth}Z)$ gives you the incoming direction of the null geodesic of that past null cone whereas ${\bar \eth} \eth Z$ identifies a point on that null geodesic. Therefore, with a very small array of observers one can identify points in the spacetime such that their null cone cuts are described by equation (\ref{nc cut}).

One can also show that $Z_{,a}$ is a null covector, namely, it satisfies

\begin{equation}\label{NSF metric}
g^{ab}Z_{,a}Z_{,b}=0.
\end{equation}
 The above equation can also be used to reconstruct the conformal metric from knowledge of $Z$. The explicit construction is given on a preferred coordinate system $(u,\omega,{\bar \omega}, R) = (Z, \eth Z, {\bar \eth}Z, {\bar \eth} \eth Z)$, and the metric coefficients are given in terms of a function
 $\Lambda(Z, \eth Z, {\bar \eth}Z, {\bar \eth} \eth Z, \zeta, {\bar \zeta})$, related to $Z$ by the equation
 $$\eth^2 Z = \Lambda.$$
 This function plays a central role in the metric reconstruction technique. If $\Lambda$ is given, to obtain a Lorentzian metric from (\ref{NSF metric}), $\Lambda$ must satisfy a set of PDEs called metricity conditions. This is the core of the Null Surface Formulation of General relativity \cite{FKN}, or NSF for short, and it gives a generalization of Cartan's work on third order ODEs and a Lorentzian metric on the solution space \cite{Cartan, Chern}. Note that if $\Lambda=0$ we obtain a flat metric and  the solution of $\eth^2 Z = 0$ is given by (\ref{flat cut}).

$\Lambda$ also has a very simple geometric meaning. Using Sachs' theorem one can show that
\begin{equation}\label{NCcuts}
\eth^2 Z=\sigma^0-\sigma_x,
\end{equation}
with $\sigma^0$ the asymptotic Bondi shear at null infinity and  $\sigma_x$ the asymptotic shear of the future null cone from $x^a$ evaluated at null infinity \cite{BKR}. In general $\sigma_x$ will always be non-vanishing for a non flat spacetime since the Weyl tensor induces shear on the future null cone from any point $x^a$. It follows from the above equation that a vanishing asymptotic shear does not correspond to a NC cut. (As a remark we point out that in the ANK approach one uses a congruence of null geodesics such that the associated asymptotic shear vanishes at null infinity.  One thus sets $\sigma_x=0$ in eq. (\ref{NCcuts}) to obtain the so called "good cut equation".)

As we are interested in describing a particular worldline whose motion will depend on $\sigma^0(u_B,\zeta, {\bar \zeta})$ we assume $\dot{\sigma}^0$ is known. Moreover, the outgoing gravitational radiation we are interested in is emitted by closed binaries, supernovae or scattering of compact sources. For those systems one can always assume they are asymptotically stationary, i.e., $\dot{\sigma}^0$ vanishes as $u_B \rightarrow -\infty$. In that limit $\sigma_I \rightarrow 0$, $\sigma^0$ is purely electric and by a supertranslation one can get rid of the electric part at that initial time. We thus assume that we work in a definite Bondi system such that $\sigma^0$ vanishes as  $u_B \rightarrow -\infty$. This restricts the Bondi supertranslation freedom to the translations of the Poincare group. Following the above results the description of the cuts in any other Bondi system will be given by $\tilde{Z}(x^a,\zeta, {\bar \zeta}) = Z(x^a,\zeta, {\bar \zeta}) +\alpha(\zeta,{\bar \zeta})$ and the $\ell= 0, 1$ parts of $Z$ and $\tilde{Z}$ do not depend on the higher harmonics of $\alpha(\zeta,{\bar \zeta})$.

Finally, we want to obtain dynamical equations for $Z$  to exhibit the explicit dependence of $\sigma_x$ on $\sigma^0$. It is clear that one cannot hope to obtain $Z$ or $\Lambda$ in closed form for an arbitrary asymptotically flat spacetime. On the other hand it is not difficult to set up a perturbation procedure off Minkowski space and obtain a first order deviation from a flat cut.

 Writing
 $$Z = Z_0+Z_1,$$
 with $Z_0$ given by (\ref{flat cut}) and
 $$\Lambda_1 = \eth^2 Z_1,$$
 one can show that $\Lambda_1$ satisfies the wave equation in Minkowski space and that it functionally depends on the Bondi shear via (see Appendix \ref{Appendix A})
 \begin{equation}\label{RNCcuts}
{\bar \eth}^2 \eth^{2}Z_1={\bar \eth}^2 \sigma^0(Z_0,\zeta,{\bar \zeta}) + \eth^2{\bar \sigma}^0(Z_0,\zeta,{\bar \zeta}).
\end{equation}
The second term in the r.h.s. of eq. (\ref{RNCcuts}) gives the relationship between $\sigma_x$ and $\sigma^0$. Since the Bondi shear is a smooth s.w. 2 function on $\scri^+$ the above equation admit regular solutions on the sphere. Thus, the first order deviation from a flat cut are smooth 2-surfaces (they can be expanded in spherical harmonics) at null infinity and are called the (linearized) regularized null cone cuts.
If $x^a(u)$ describes a worldline in Minkowski space, the function $Z(x^a(u),\zeta, {\bar \zeta})$ describes a one parameter family of cuts. To show  that this family is NU we perform a Taylor expansion
$$Z(x^a(u+\delta u),\zeta, {\bar \zeta})= Z(x^a(u),\zeta, {\bar \zeta})+ v^a\partial_a Z \delta u,$$
where $v^a \equiv \partial_u x^a$ and $\delta u> 0$. If we assume $v^a$ is future pointing with respect to the flat metric, it then follows that
$$ v^a\partial_a Z> 0,$$ since $Z^a$ is null and future pointing (for the flat metric) for small values of $\sigma_B$. We conclude that this monoparametric family never intersects itself and it is a well behaved NU foliation.

Solving for (\ref{RNCcuts}) yields
\begin{equation}\label{Z-1}
Z=R^{0}-\frac{1}{2}R^{i}Y_{1i}^{0}+\left( \frac{\sigma _{R}^{ij}}{12}+\frac{\sqrt{2}}{72}\dot{\sigma}_{I}^{ig}R^{f}\epsilon ^{gfj}\right) Y_{2ij}^{0},
\end{equation}
with $Y_0^0,Y^0_{1i}$ and $Y_{2ij}^{0}$ the tensorial spin-s harmonic expansion \cite{ngilb}.
Note that $Z$ depends on the real and imaginary part of the Bondi shear \cite{KQ,BKR}. This is what one would expect in a perturbation expansion since the imaginary part of the Bondi shear is related to the current quadrupole moment whereas the real part comes from the mass quadrupole moment\cite{Kidder}.

The first order solution (\ref{Z-1}) will be used to define center of mass and spin for isolated sources of gravitational radiation. It will also be used to compare our results with those derived in the ANK and postNewtonian formulations.

Finally, it is a fair question to ask what happens to the above construction if one goes  beyond the linearized approximation.

If we assume the spacetime is Ricci flat in a neighborhood of $\scri ^{+}$ one obtains the field equation for $Z$ \cite{FKN} (see appendix \ref{Appendix A}). The field equation exhibits the non Huygens nature of the NC cuts showing explicitly which term is responsible for caustics. Thus, a generic NC cut is not a smooth 2-surface at null infinity. However, if $\dot{\sigma}^0$ is small both in the past or in the future  of some small interval of time, one expects that the leading contribution to the solution comes from the Huygens part of field equation,
\begin{equation}\label{RNC-cuts}
{\bar \eth}^2 \eth^{2}Z={\bar \eth}^2 \sigma^0(Z,\zeta,{\bar \zeta}) + \eth^2{\bar \sigma}^0(Z,\zeta,{\bar \zeta}),
\end{equation}
referred to as the Regularized Null Cone cut equation or RNC cut equation for short. Since (\ref{RNC-cuts}) only contains $l>2$ terms in a spherical harmonic decomposition, the  kernel of (\ref{RNC-cuts}) is a 4 dim space $x^a$, i.e. a flat cut $Z_0= x^a\ell_a$.

Equation (\ref{RNC-cuts}) or its linearized version (\ref{RNCcuts}) should be compared with the good cut equation
\begin{equation}\label{good cut}
\eth^2 Z_C= \sigma^0(Z_C,\zeta,{\bar \zeta}),
\end{equation}

Note that the good cut equation yields complex cuts with vanishing shear whereas the NRC cut equation yields NU cuts whose shear depends linearly on the Bondi shear.

Thus, from the point of view of available structures at null infinity we could start with the RNC cut equation (\ref{RNC-cuts}). On its 4-dim solution space one constructs a Lorentzian metric following the NSF procedure\cite{FKN}. A perturbative solution gives a Minkowski space together with flat cuts (\ref{flat cut}) at its lowest order and the linearized RNC cuts (\ref{Z-1}) at first order.

\section{Linkages and the angular momentum-center of Mass Tensor}
For axially symetric spacetimes the Komar integral constructed from the axial Killing field yields a natural definition of angular momentum that it is a conserved quantity in vacuum and has a flux law in Einstein-Maxwell spacetimes \cite{KQ}. This idea can be generalized to asymptotically flat spacetimes by first introducing the notion of asymptotic Killing vectors and then giving a  generalization of the Komar integral, the Winicour-Tamburino linkage \cite{TW}, which yields the Komar formula when the spacetime has a Killing symmetry. We will use in future sections these concepts to define the spin, total angular momentum and center of mass of an asymptotically flat spacetime.
\subsection{The Asymptotic Symmetry Group}
First we introduce the  generators $\xi^a$ of asymptotic symmetries on a neighborhood of $\scri^+$ as smooth solutions of the asymptotic Killing equation \cite{GW}
\begin{eqnarray}
\xi _{a;b}+\xi _{b;a} &=&O(r^{-n}) \label{killing} \\
(\xi _{a;b}+\xi _{b;a})l^{b\ast } &=&0. \label{killing2}
\end{eqnarray}

Here $l^{b\ast}$ is a null vector tangent to the generators of each outgoing null hypersurface in $\scrm$ and $n$ differs with the choice of components \cite{nu}. The second equation represents the Killing propagation law along the null hypersurface \cite{Wini}. At $\scri^+$ the collection of all solutions form the BMS algebra $\scrl$ \cite{sachs}. If $\xi^a \propto n^a$ they define the supertranslation subalgebra $\scrt$ and the quotient $\scrl / \scrt$ is isomorphic to the Lorentz group \cite{GW}. This subalgebra is realizad by an equivalence class $[\xi^a]$ where $\xi^a \sim \xi^{\prime a}$ if $\xi^a-\xi^{\prime a} \propto n^{a\ast}$.
Equations (\ref{killing}) and (\ref{killing2}) can be solved by direct integration using the spin-coefficient \cite{LMN}. The results may be written as
\begin{eqnarray}
\xi ^{a}&=&A l^{a\ast }+Bn^{a\ast }+C \bar{m}^{a\ast}+\bar{C}m^{a\ast }
\end{eqnarray}
where
\begin{eqnarray*}
A &=&A_{1}r+A_{0}+A_{-1}r^{-1}+O(r^{-2}) \\
B &=&B_{0} \\
C &=&C_{1}r+C_{0}+C_{-1}r^{-1}+O(r^{-2})
\end{eqnarray*}
and
\begin{eqnarray*}
&&A_{1} =-(1/Z^{\prime })(B_{0}Z^{\prime })^{\prime }, \\
&&A_{0} =\eth ^{\ast }\bar{\eth }^{\ast }B_0+B_{0}\eth ^{\ast }\bar{\eth }%
^{\ast }\ln P, \\
&&A_{-1}=\frac{1}{2}[B_{0}(\psi _{2}^{0\ast }+\bar{\psi}_{2}^{0\ast })+\bar{%
C}_{1}\psi _{1}^{0\ast }+C_{1}\bar{\psi}_{1}^{0\ast }], \\
&&C_{1} =a(\zeta ,\bar{\zeta})/Z^{\prime }, \quad with \quad \eth a =0, \\
&&C_{0} =\eth ^{\ast }B_{0}+\bar{C}_{1}\sigma ^{0\ast }, \\
&&C_{-1} =0, \\
&&B_{0} =b(\zeta ,\bar{\zeta})/Z^{\prime }-(1/2Z^{\prime
})\int_{0}^{u}Z^{\prime 3}[\eth (\bar{a}Z^{\prime -2})+\bar{\eth }%
(aZ^{\prime -2})]du.
\end{eqnarray*}
Note that the only freedom is in $b(\zeta,\bar\zeta)$, the supertranslation freedom, and solutions to $\eth a=0$, which correspond to the homogeneous Lorentz transformation.
\subsection{Linkages in Asymptotically Flat Spacetimes}
Given a $u=const.$ null foliation, which can be either NU or Bondi, introducing an affine parameter $r$ and constructing the $r=const.$ 2-surface with surface element $l^{\ast [a}\hat{n}^{\ast b]}dS$, the linkage integral is defined as \cite{Wini2}

\begin{equation}\label{linkage}
L_{\xi }(\scri^+)=-\frac{1}{16\pi}\underset{r\rightarrow \infty }{lim}\int \left( \nabla
^{\lbrack a}\xi ^{b]}+\nabla _{c}\xi ^{c}l^{\ast [a}\hat{n}^{\ast
b]}\right) l_{a}^{\ast }\hat{n}_{b}^{\ast}dS,
\end{equation}
Note that $\hat{n}^{\ast b}$ is not one of the associated null vectors of the NU tetrad. Whereas $n^{\ast b}$ is parallel propagated along $l^{\ast a}$, $\hat{n}^{\ast b}$ is orthogonal to the $u=const.$, $r=const.$ surface. It can be rewritten in terms of the NU tetrad via a null rotation around $l^{\ast b}$ as \cite{LMN}
\begin{equation}
\hat{n}^{\ast b}=n^{\ast b}-\bar{\omega}^{\ast}m^{\ast b}-\omega^{\ast}\bar{m}^{\ast b}+\omega\bar{\omega}l^{\ast b}
\end{equation}
with
\begin{equation}
\omega^\ast=-(\bar{\eth}^{\ast}\sigma^{0\ast})r^{-1}+O(r^{-2})
\end{equation}

This scalar linear functional of  $\xi ^{b}$ transforms as an adjoint representation of the BMS group. If $\xi ^{b}$ is a translation eq. (\ref{linkage}) yields the Bondi energy momentum vector. Likewise, if $\xi ^{b}$ belongs to the Lorentz subgroup, eq. (\ref{linkage}) can be used to define the notion of the mass dipole and angular momentum. Solving the asymptotic Killing equation by making use of the radial dependence of the spin coefficients and tetrad components, one can show that this linkage integral can be written as \cite{LMN},

\begin{eqnarray*}
L&=&\frac{1}{8 \pi \sqrt{2}} Re\int [ b\left( \frac{\psi _{2}^{0\ast}+\sigma^{0\ast}\lambda ^{0\ast}-\eth^{\ast 2}\bar{\sigma}^{0\ast}}{Z^{\prime 3}}\right)\\
&&+\bar{a} \left( \frac{2\psi _{1}^{0\ast}-2\sigma ^{0\ast}\eth^{\ast}\bar{\sigma}^{0\ast}-\eth^{\ast}(\sigma^{0\ast}\bar{\sigma}^{0\ast})}{Z^{\prime 3}}\right)] dS
\end{eqnarray*}

putting $b=0$ we obtain

\begin{eqnarray}\label{link}
L_{DJ} = \frac{Re}{8 \pi \sqrt{2}}\int \bar a \left[ \frac{2\psi _{1}^{0\ast}-2\sigma^{0\ast}\eth^{\ast} \bar{\sigma}^{0\ast}-\eth^{\ast}(\sigma ^{0\ast}\bar{\sigma}^{0\ast})}{Z^{\prime 3}}\right]dS \nonumber\\
\end{eqnarray}
where $\bar a=\bar{a}^i Y^{-1}_{1i}$ with $\bar{a}^i$ three complex constants. The three complex, i.e six real, values of (\ref{link}) are by definition, the components of the mass dipole - angular momentum tensor (for more details the reader can see ref. \cite{LMN}). To obtain those components it is quite convenient to define a complex vector  $D^{\ast}_i+\frac{\mathtt{i}}{c}J^{\ast}_i$ where $i$ symbolize the vectors (1,0,0), (0,1,0) and (0,0,1) as
\begin{eqnarray}\label{link2}
D^{\ast}_i+\frac{\mathtt{i}}{c}J^{\ast}_i = \int Y^{-1}_{1i} \left[ \frac{2\psi _{1}^{0\ast}-2\sigma^{0\ast}\eth^{\ast} \bar{\sigma}^{0\ast}-\eth^{\ast}(\sigma ^{0\ast}\bar{\sigma}^{0\ast})}{8 \pi \sqrt{2}Z^{\prime 3}}\right]dS \nonumber\\
\end{eqnarray}

It is worth mentioning that at a linearized level and for stationary spacetimes, the real and imaginary parts of $\psi _{1}^{0}$ capture the notion of the two form that defines the center of mass and angular momentum and transform appropriately under the Lorentz transformation. The linkage is a natural generalization for asymptotically flat spacetimes.

It is also worth mentioning that the value of the linkage depends on the choice of section introduced for its definition \cite{GW}. This is analogous to the freedom in special relativity with the choice of origin for the definition for center of mass or angular momentum. The main difference is that whereas in relativistic mechanics the freedom is a point on the spacetime, in the definition of a linkage the freedom is a whole section, an infinite set of constants, one for each coefficient in a spherical harmonic decomposition. Consequently, if one now has a NU foliation, where each coefficient now depends on the Bondi time, the freedom becomes an infinite set of functions of time a priori without physical meaning.

However, in what follows below, we will restrict this infinite freedom to four functions that describe a worldline in the solution space of the RNC cuts. From its geometrical meaning there is a one to one correspondence between worldlines in the solution space and a  RNC foliation at null infinity. Furthermore, by defining the notion of mass dipole moment and requiring that for one worldline of the RNC foliation the mass dipole moment vanishes, one gets the right number of equations from which a special worldline is found. This special worldline will be called the center of mass worldline. Finally, restricting the angular momentum to this special RNC cut yields the notion of spin or intrinsic angular momentum.


\section{Main Results}

\subsection{Definitions of Center of Mass and Angular Momentum}
\quad Directly from (\ref{link2}) we define the mass dipole moment and angular momentum associated with a RNC foliation as
\begin{eqnarray}\label{DJ}
D^{\ast}_i+\frac{\mathtt{i}}{c}J^{\ast}_i=\frac{-c^{2}G^{-1}}{12\sqrt{2}}\left[ \frac{2\psi _{1}^{0\ast}-2\sigma
^{0\ast}\eth^{\ast}\bar{\sigma}^{0\ast}-\eth^{\ast}(\sigma ^{0\ast}\bar{\sigma}^{0\ast})}{Z^{\prime 3}}\right]^{i}. \nonumber\\
\end{eqnarray}

The six functions of the NU time $u$ defined above functionally depend on the particular worldline $x^a(u)$ that characterizes each RNC cut.

We then impose a condition on a special RNC foliation, i.e., on a special worldline (\ref{Z-1}), at each $u=const.$ cut,  the mass dipole moment $D^{\ast i}$ vanishes. This condition is given by
\begin{equation}
Re\left[ \frac{2\psi _{1}^{0\ast}-2\sigma^{0\ast}\eth^{\ast} \bar{\sigma}^{0\ast}-\eth^{\ast}(\sigma ^{0\ast}\bar{\sigma}^{0\ast})}{Z^{\prime 3}}\right]^{i}=0. \label{CoM}
\end{equation}
By adequately choosing $x^a(u)$ one has enough freedom to satisfy the above equation for each value of $u$. Since the 4-velocity of the worldline is normalized to one (using the spacetime metric), we use this norm to fix the timelike component of the worldline coordinate. Thus, the freedom left are the spatial components of the worldline $x^a(u)$ and the above equation gives three algebraic equations from which these components are obtained. This special worldline will be called the center of mass worldline. The angular momentum $J^{i\ast}$ evaluated at the center of mass will be called intrinsic angular momentum $S^i$, i.e.,

\begin{eqnarray}
S^{i}=-\frac{c^{3}}{12\sqrt{2}G}Im\left[ \frac{2\psi _{1}^{0\ast}-2\sigma^{0\ast}\eth^{\ast} \bar{\sigma}^{0\ast}-\eth^{\ast}(\sigma ^{0\ast}\bar{\sigma}^{0\ast})}{Z^{\prime 3}}\right]^{i}. \nonumber\\ \label{Si}
\end{eqnarray}

The above equations have been obtained from two surface integrals on a particular RNC cut foliation, namely, the center of mass foliation. Thus, they have a well defined geometrical meaning. We now solve eq. (\ref{CoM}) explicitly on a Bondi frame since variables like gravitational radiation, mass loss, linear momentum, are easier to define in Bondi coordinates. To write down the mass dipole moment and and angular momentum (\ref{DJ}) in Bondi coordinates it is convenient to define analogous quantities in a Bondi tetrad, i.e.,
\begin{equation}\label{DJB}
D^{i}+\mathtt{i}c^{-1}J^{i}=-\frac{c^{2}}{12\sqrt{2}G}\left[ 2\psi _{1}^{0}-2\sigma
^{0}\eth \bar{\sigma}^{0}-\eth(\sigma ^{0}\bar{\sigma}^{0})\right]^{i}.
\end{equation}
Using the relations between the NU and the Bondi null vectors given by (\ref{la}-\ref{sigma*}) to transform the quantities $(\psi _{1}^{0\ast }, \sigma^{0\ast}, \eth^{\ast})$$\rightarrow$$(\psi _{1}^{0}, \sigma^{0}, \eth)$, one can write (\ref{DJ}) as
\begin{eqnarray}
D^{\ast i}(u) &=&D^{i}(u_B)+\frac{3c^{2}}{6\sqrt{2}G}Re[\eth Z(\Psi -\eth ^{2}\bar{\sigma}^{0})+F]^{i}\label{Dexp} \\
J^{i\ast }(u) &=&J^{i}(u_B) +\frac{3c^{3}}{6\sqrt{2}G}Im[\eth Z(\Psi -\eth ^{2}\bar{\sigma}^{0})+F]^{i}\label{Jexp}
\end{eqnarray}
with
\begin{eqnarray}\label{F}
F&=&-\frac{1}{2}(\sigma ^{0}\eth \bar{\eth }^{2}Z+\eth ^{2}Z\eth \bar{\sigma}^{0}-\eth ^{2}Z\eth \bar{\eth }^{2}Z)\nonumber\\
&&-\frac{1}{6}(\bar{\sigma}^{0}\eth ^{3}Z+\bar{\eth }^{2}Z\eth \sigma ^{0}-\bar{\eth }^{2}Z\eth ^{3}Z).
\end{eqnarray}
If we insert the center of mass RNC cut $Z_1$ in (\ref{Dexp}),  then its l.h.s. vanishes on a $u=$ const. surface and we obtain an algebraic equation to be solved for $R^{i}(u)$. Equation (\ref{Jexp}) then gives a relationship between $S^i$ and $J^{i}$, the intrinsic and total angular momentum respectively.
\subsection{Approximations and assumptions} \label{supuestos}
Although the main equations have been presented above, to obtain the explicit form of the worldline in this work we will make the following assumptions.
\begin{itemize}
  \item $\sigma=0$ for some initial Bondi time, usually taken to be $-\infty$.
  \item $R^i$ is a small deviation form the coordinates origin.
  \item $R^0=u$ assuming the slow motion approximation.
  \item The Bondi shear only has a quadrupole term.
\end{itemize}
The first assumption fixes the supertranslation freedom and is consistent with our choice of null cone cut, namely, the freedom in the solution of eq. (\ref{RNCcuts}) is only a translation between two Bondi frames. The second assumption is a working simplification. Since we are particularly interested in the acceleration of the center of mass, which is quadratic in the gravitational radiation, we want to ignore terms like $R^2 \sigma$. Finally, the third assumption is a physical one. Since in most astrophysical processes less than $25\%$ of the total mass is lost as gravitational radiation the gamma factor for the center of mass velocity is about $1.003$. Putting it in other words, even if two coalescing stars are approaching each other at relativistic speeds, if the center of mass is initially at rest it will never acquire a relativistic velocity.

In principle all of these assumptions can be relaxed but since we want to make direct comparisons with other formulations, like the ANK approach or the PN equations of motion, they are needed for these purposes.The ANK approach uses the same Bondi gauge as ours whereas the PN formulation selects an initial time where the system is stationary and the metric is flat.

It is possible to extract several important formulae relating the dynamical evolution of mass, momentum, etc. by expanding the Bianchi identities in a spherical harmonics decomposition. Using the tensorial spin-s spherical harmonics \cite{ngilb}; $Y^0_0, Y^0_{1i}, Y^0_{2ij}$, etc., one can expand the relevant scalars at null infinity as
\begin{eqnarray} \label{expan_tensorial}
\sigma^0 &=&\sigma ^{ij}(u_{B})Y_{2ij}^{2}(\zeta,\bar \zeta ), \nonumber\\
\psi _{1}^{0} &=&\psi _{1}^{0i}(u_{B})Y_{1i}^{1}(\zeta,\bar \zeta )+\psi_{1}^{0ij}(u_{B})Y_{2ij}^{1}(\zeta,\bar \zeta ), \\
\Psi &=&-\frac{2\sqrt{2}G}{c^{2}}M-\frac{6G}{c^{3}}P^{i}Y_{1i}^{0}(\zeta,\bar \zeta )+\Psi ^{ij}(u_{B})Y_{2ij}^{0}(\zeta,\bar \zeta )\nonumber,
\end{eqnarray}
Note that the complex tensor $\sigma^{ij}$ represents the quadrupole momentum of the gravitational wave.

Now, from eq. (\ref{RNCcuts}) if we write $x^a(u)$ as $(R^{0}(u),R^{i}(u))$, assuming the Bondi shear only has a quadrupole term, and using the tensorial spin-s harmonic expansion, this solution is given as
\begin{equation}\label{linear worldline}
Z_1(u,\zeta,\bar \zeta)= R^{0}(u)-\frac{1}{2}R^{i}(u)Y_{1i}^{0}+\frac{1}{12}\sigma_R^{ij}(u)Y_{2ij}^{0}.
\end{equation}
the freedom left in (\ref{linear worldline}) is an arbitrary worldline in a fiducial spacetime. Choosing $u$ as the proper time, we can easily solve for $R^0(u)$ in terms of the spatial components of the 4-velocity. Furthermore, in the slow motion approximation $R^0(u)=u+O(v^2)$.

\subsection{The center of mass and spin} \label{CoM Spin}
The center of mass worldline $R^{i}(u)$ is obtained from (\ref{Dexp}) by demanding that the l.h.s. vanishes on the $u=const.$ cut when $u_B=Z_1(u,\zeta,\bar \zeta)$ is inserted in the r.h.s. of the equation. Furthermore, since by assumption $R^{i}(u)$ and $\sigma_R^{ij}(u)$ are small, we write
\begin{equation}\label{Z1}
Z_1=u+\delta u=u -\frac{1}{2}R^{i}(u)Y_{1i}^{0}+\frac{1}{12}\sigma _{R}^{ij}(u)Y_{2ij}^{0},
\end{equation}
and make a Taylor expansion of the Bondi tetrad variables up to first order in $\delta u$. We write (\ref{Dexp}) as
\begin{eqnarray}
0 &=&D^{i}(u+\delta u)+\frac{3c^{2}}{6\sqrt{2}G}Re[(\Psi -\eth ^{2}\bar{\sigma}^{0})\eth \delta u+F]^{i} \nonumber \\
&=&D^{i}(u)+[\dot{D}(u)\delta u]^{i}+\frac{3c^{2}}{6\sqrt{2}G}Re[(\Psi -\eth ^{2}\bar{\sigma}^{0})\eth \delta u +F]^{i} \nonumber \\
&=&D^{i}(u)+\frac{c^{2}}{6\sqrt{2}G}Re[(\eth \Psi -\eth ^{3}\bar{\sigma}^{0})\delta u]^{i}\nonumber\\
&&+\frac{3c^{2}}{6\sqrt{2}G}Re[(\Psi -\eth ^{2}\bar{\sigma}^{0})\eth \delta u+F]^{i}, \nonumber\\
\end{eqnarray}
where we have used eq. ({\ref{psi1prima}}) to rewrite $\dot D^i$. Note that in this case the second line in the definition of $F$ (\ref{F}) vanishes since $\sigma ^{0}$ only has quadrupole terms. In the remaining terms  of (\ref{Dexp}) we simply replace $u_B$ by $u$ as the extra terms are cubic or higher in the expansion variables. Solving for $R^{i}$ from the above equation yields
\begin{equation}\label{momdip}
MR^{i}=D^{i}+\frac{8}{5\sqrt{2}c}\sigma _{R}^{ij}P^{j},
\end{equation}
where $\sigma _{R}^{ij}$ and $\sigma _I^{ij}$ are respectively the real and the imaginary part of $\sigma^{ij}$. Note that inserting eq. (\ref{momdip}) in (\ref{Z1}) yields
\begin{equation}\label{ZCM}
Z_{CM}=u -\frac{1}{2M}(D^{i}+\frac{8}{5\sqrt{2}c}\sigma _{R}^{ij}P^{j})Y_{1i}^{0}+\frac{1}{12}\sigma _{R}^{ij}Y_{2ij}^{0},
\end{equation}
the special NU foliation that represents the center of mass worldline.

As it was mentioned previously, replacing eq. (\ref{momdip}) in the imaginary part of (\ref{DJ}) yields the spin of the system. To do that we start with the relationship (\ref{Jexp}) \begin{equation*}
J^{\ast i}(u)=J^{i}(u+\delta u)+\frac{3c^{3}}{6\sqrt{2}G}Im[(\Psi -\eth ^{2}\bar{\sigma}^{0})\eth \delta u+F]^{i},
\end{equation*}
perform a Taylor expansion,
\begin{equation*}
J^{\ast i}(u)=J^{i}(u)+[\dot{J}(u)\delta u]^{i}+\frac{3c^{3}}{6\sqrt{2}G}Im[(\Psi -\eth ^{2}\bar{\sigma}^{0})\eth \delta u +F]^{i},
\end{equation*}
 and use the Bianchi identities,
\begin{eqnarray}
J^{\ast i}&=&J^{i}(u)+\frac{c^{3}}{6\sqrt{2}G}Im[(\eth \Psi -\eth ^{3}\bar{\sigma}^{0})\delta u]^{i}\\
&&+\frac{3c^{3}}{6\sqrt{2}G}Im[(\Psi -\eth ^{2}\bar{\sigma}^{0})\eth \delta u+F]^{i}. \nonumber
\end{eqnarray}
Finally, using eq. (\ref{ZCM}) gives

\begin{equation}\label{angmomentum}
S^{i}=J^{i}-R^{j}P^{k}\epsilon ^{ijk}.
\end{equation}
Note that this equation is exactly the same formula as in Newtonian theory although no post-Newtonian approximation has been assumed.

\subsection{Dynamical Evolution} \label{DynEvo}
The time evolution of $D^i$ and $J^i$ follows from the Bianchi identity for $\psi_1^0$, where we must insert the proper factor of $\sqrt{2}$ to account for the retarded Bondi time, i.e., the retarded time, $u_{ret} = \sqrt{2} u_B$. The use of the
retarded time, $u_{ret}$, is important in order to obtain the correct numerical factors in the expressions for the final physical results \cite{ANK}. Note that the two last important eqs. (\ref{momdip}) and (\ref{angmomentum}) remain unchanged in term of $u_{ret}$ or $u_B$. However, for the rest of the paper, we adopt  the symbol ``dot'' for $\partial_{u_{ret}}$.

Then, we use the definition (\ref{DJB}) and replace the real and imaginary $l=1$ component of (\ref{psi1prima}) to obtain
\begin{eqnarray}
\dot{D}^{i}&=&P^{i},\label{realpart}\\
\dot{J}^{i}&=&\frac{c^{3}}{5G}(\sigma _{R}^{kl}\dot{\sigma}_{R}^{jl}+\sigma _{I}^{kl}\dot{\sigma}_{I}^{jl})\epsilon ^{ijk}.\label{impart}
\end{eqnarray}
In the same way taking the $l=0, 1$ part of (\ref{psiprima}) yields the mass loss equation and the linear momentum time rate, namely,
\begin{eqnarray}
\dot{M}&=&-\frac{c}{10G}(\dot{\sigma}_{R}^{ij}\dot{\sigma}_{R}^{ij}+\dot{\sigma}_{I}^{ij}\dot{\sigma}_{I}^{ij}),\label{mpunto}\\
\dot{P}^{i} &=&\frac{2c^{2}}{15G}\dot{\sigma}_{R}^{jl}\dot{\sigma}%
_{I}^{kl}\epsilon ^{ijk}. \label{ppunto}
\end{eqnarray}
Note that in this Bondi gauge $\sigma _{R}^{ij}=h_+^{ij}$ and $\sigma _{I}^{ij}=h_\times^{ij}$ strains in the transverse traceless gauge \cite{ttgauge}.
Now, taking a time derivative of eq. (\ref{momdip}), using eq. (\ref{realpart}), and writing up to quadratic terms in $\sigma^{ij}$, gives
\begin{equation}\label{momento}
M\dot{R}^{i}=P^{i}+\frac{8}{5\sqrt{2}c}\dot{\sigma}_{R}^{ij}P^{j},
\end{equation}
the relationship between the velocity of the center of mass $\dot{R}^i$ and the Bondi momentum. It departs from the Newtonian formula by radiation terms.

Finally, taking one more Bondi time derivative of (\ref{momento}) yields the equation of motion for the center of mass,
\begin{equation}
M\ddot{R}^{i}=\frac{2c^{2}}{15G}\dot{\sigma}_{R}^{jl}\dot{\sigma}%
_{I}^{kl}\epsilon ^{ijk}+\frac{8}{5\sqrt{2}c}\ddot{\sigma}_{R}^{ij}P^{j}.
\end{equation}
The r.h.s. of the equation only depends on the gravitacional data at null infinity and the initial mass of the system.

Similarly, taking a time derivative of (\ref{angmomentum}) together with (\ref{impart}) gives
\begin{equation}\label{Sdot}
\dot{S}^{i}=\dot{J}^{i}=\frac{c^{3}}{5G}(\sigma _{R}^{kl}\dot{\sigma}_{R}^{jl}+\sigma _{I}^{kl}\dot{\sigma}_{I}^{jl})\epsilon ^{ijk}.
\end{equation}
This equality is also true in Newtonian mechanics for an isolated system (with both terms being equal to zero). However, in GR the angular momentum of an isolated system is not conserved since it is being carried away by the gravitational radiation.
\subsection{Comparison with ANK equations}
\quad In this subsection we compare the (ANK) equations of motion with the ones obtained in our approach. Before that we list the main differences between the approaches,
\begin{enumerate}
  \item We give a definition of angular and mass dipole momenta based on TWG linkages, the ANK uses the $\ell=1$ part of $\psi_1^0$ for these definitions.
  \item The ANK approach relies on asymptotically vanishing shears, this approach uses non vanishing shears obtained from the RNC cut equation.
  \item The solution space of the good cut equation is complex manifold, the solution space of the RNC cut equation is real.
  \item The ANK approach defines the intrinsic angular momentum as the imaginary part of a complex worldline. We evaluate the angular momentum on the center of mass to define the spin.
\end{enumerate}

Thus, it is interesting to see if the final equations in these two formulations have some similarities. To proceed with the comparison we identify the flat metric of our construction with the real flat metric used in the ANK approach to write the equations of motion for the center of mass worldline.

It is also important to note that the Bondi mass $M$ and the linear momentum $P^i$ have the same definition in both approaches. First we introduce the mass dipole moment, angular momentum and spin definitions given in the ANK formalism \cite{ANK}

\begin{eqnarray}
D_{\text{\tiny\emph{ANK}}}^{i}&=&-\frac{c^{2}}{6\sqrt{2}G}\psi _{1R}^{0i},\\
J_{\text{\tiny\emph{ANK}}}^{i}&=&-\frac{c^{3}}{6\sqrt{2}G}\psi _{1I}^{0i},\\
S_{\text{\tiny\emph{ANK}}}^{i}&=&cM\xi _{I}^{i}.
\end{eqnarray}
Now computing the component $l=1$ of eq. (\ref{DJB}) we can write
\begin{eqnarray}
D^{i}&=&-\frac{c^{2}}{6\sqrt{2}G}\psi _{1R}^{0i}+\frac{c^{2}}{5G}\sigma
_{R}^{jl}\sigma _{I}^{kl}\epsilon ^{ijk} +  \mbox{\small higher harmonics}\nonumber\\
J^{i}&=&-\frac{c^{3}}{6\sqrt{2}G}[\psi _{1}^{0}-\sigma^{0}\eth \bar{\sigma}^{0}-\frac{1}{2}\eth(\sigma ^{0}\bar{\sigma}^{0})]_{I}^{i}.
\end{eqnarray}
The relationship between the mass dipole moment and angular momentum with the asymptotic fields at null infinity are different in both formalisms.  These differences are a consequence of the definitions used in both formulations. Whereas in our approach we integrate a two form with values on the BMS algebra, in the ANK approach one directly uses $\psi _{1}^{0i}$ for the definitions.

The angular momenta in the ANK formulation is only defined for quadrupole radiation, where most of the definitions available in the literature agree. However, one could forsee potential problems for $J^i_{\text{\tiny\emph{ANK}}}$ if one considers higher multipole moments in the radiation data and/or spacetimes with symmetry. The fact that $\psi _{1I}^{0i}$ is not conserved for axially symmetric spacetimes is a clear indication that the ANK definition must be changed when including higher multipole moments \cite{KQ}. It is worth mentioning that only for quadrupole radiation both formulae agree. We obtain non vanishing extra terms when octupole data is included (see appendix \ref{Appendix B}).

When comparing the relationship between the center of mas worldline, and spin and the geometrical quantities at null infinity like the Bondi mass, momentum, etc., we will only consider quadrupole radiation data.

In the ANK approach one has \cite{ANK}
\begin{eqnarray}
P^{i} &=&M\dot{\xi}_{R}^{i}+\frac{4}{3c\sqrt{2}}\dot{\sigma}_{R}^{ij}P^j+\frac{c^{2}}{G}(\sigma _{R}^{jl}\sigma _{I}^{kl}\dot{)}\epsilon ^{ijk},
\end{eqnarray}
where $\xi_{R}^{i}$ is the center of mass worldline. In our formulation, from eq. (\ref{momento}) we get
\begin{equation}
P^{i}=M\dot{R}^{i}-\frac{8}{5\sqrt{2}c}\dot{\sigma}_{R}^{ij}P^{j},
\end{equation}

The main difference between the above equations is the last term in the ANK, which is missing in our equation. Note also a different factor with an opposite sign in front of the second term. This difference can be traced back in the ANK formulation to the use of the relation $\Psi^{ij}=-\bar{\sigma}^{ij}$ in eq. (6.33) \cite{ANK}. However, this relationship contradicts eq. (\ref{psiprima}) as one can see by deriving the relationship with respect to time and getting $\dot\Psi^{ij}=-\dot{\bar{\sigma}}^{ij}$. It is clear from eq. (\ref{psiprima}) that $\dot{\Psi}^{ij}$ must be quadratic in $\dot{\sigma}^{ij}$ . Thus, some derivations in the ANK formulation, and in particular the above relation, are incorrect.

The ANK equations for the angular momentum is given by
\begin{equation}
J^{i}_{\text{\tiny\emph{ANK}}}=S_{\text{\tiny\emph{ANK}}}^{i}+\xi _{R}^{j}P^{k}\epsilon ^{ijk}+\frac{4}{5\sqrt{2}}
P^{k}\sigma _{I}^{ki}.
\end{equation}
whereas we obtain
\begin{equation}
J^{i}=S^{i}+R^{j}P^{k}\epsilon ^{ijk}.
\end{equation}
Another subtle but important difference is that our definition of spin is via a linkage formulation whereas in the ANK formulation the spin is an intrinsic property of a complex worldline, by definition it is the imaginary part of a complex worldline.

Finally in the ANK formalism the equation of motion for the center of mass is given by
\begin{equation}
M\ddot{\xi}_{R}^{i} =\frac{2\sqrt{2}c^{2}}{15G}%
\dot{\sigma}_{R}^{jl}\dot{\sigma}_{I}^{kl}\epsilon ^{ijk}-\frac{c^{2}}{G}(\sigma _{R}^{jl}\sigma _{I}^{kl}\ddot{)}\epsilon ^{ijk}-\frac{4}{3c\sqrt{2}}\ddot{\sigma}_{R}^{ij}P^j,
\end{equation}

while in our formalism it is given by
\begin{equation}
M\ddot{R}^{i}=\frac{2c^{2}}{15G}\dot{\sigma}_{R}^{jl}\dot{\sigma}%
_{I}^{kl}\epsilon ^{ijk}+\frac{8}{5\sqrt{2}c}\ddot{\sigma}_{R}^{ij}P^{j}.
\end{equation}
Although both formulations agree for stationary spacetimes, they differ when
gravitational radiation is present.

\subsection{Comparison with PN equations}
In this subsection we partially compare the evolution equations obtained in our approach with those coming from the PN formalism. In principle, a full comparison between these approaches can be a formidable task, i.e., the PN start with definitions in the near zone with multipoles defined in terms of the source whereas the asymptotic formulation defines radiative multipole moments. The asymptotic formulation has exact equations of motion for mass, momentum and angular momentum whereas in the PN approach one builds up the loss due to gravitational radiation valid up to the level of approximation considered since apriori one does not have available an exact formula. Nevertheless it is very useful to try to build a bridge between these approaches and see whether or not they yield equivalent equations of motion for a compact source emitting gravitational radiation.

We compare below the evolution equations for the total mass, momentum and angular momentum of a compact source of gravitational radiation. In both formalisms, a dot derivative means derivation with respect with the retarded time, as one can see following ref.\cite{Blanchet} page 6 and 27.

In the PN equations the radiative energy loss, the linear and angular momentum loss are given by (in units of $G=c=1$) \cite{GoBa,BlanchetQ}
\begin{eqnarray}
\dot{E}_{\text{{\tiny \emph{PN}}}} &=&-\frac{1}{5}\dot{U}^{ij}\dot{U}^{ij}-%
\frac{16}{45}\dot{V}^{ij}\dot{V}^{ij}-\frac{1}{189}\dot{U}^{ijk}\dot{U}%
^{ijk}\nonumber\\
&&-\frac{1}{84}\dot{V}^{ijk}\dot{V}^{ijk} \\
\dot{P}_{\text{{\tiny \emph{PN}}}}^{i} &=&\left( \frac{16}{45}\dot{U}^{kl}%
\dot{V}^{jl}+\frac{1}{126}\dot{U}^{klm}\dot{V}^{jlm}\right) \epsilon ^{ijk} \nonumber\\
&&-\frac{2}{63}(\dot{U}^{jk}\dot{U}^{ijk}+2\dot{V}^{jk}\dot{V}^{ijk}) \\
\dot{J}_{\text{{\tiny \emph{PN}}}}^{i} &=&-\left( \frac{2}{5}U^{kl}\dot{U}%
^{jl}+\frac{32}{45}V^{kl}\dot{V}^{jl}\right)\epsilon ^{ijk}\nonumber\\
&&-\left(\frac{1}{63}U^{klm}\dot{U}^{jlm}+\frac{1}{28}V^{klm}\dot{V}^{jlm}\right)\epsilon ^{ijk}
\end{eqnarray}

where in the above equations the quadrupole as well as octupole terms have been included.

To compare both approaches, we must include in our formalism the octupole contribution to the equations for the mass, angular and linear momentum (see Appendix \ref{Appendix B}). In this way we can write our equations (in term of $G=c=1$) as
\begin{eqnarray}
\dot{M} &=&-\frac{1}{10}(\dot{\sigma}_{R}^{ij}\dot{\sigma}_{R}^{ij}+\dot{%
\sigma}_{I}^{ij}\dot{\sigma}_{I}^{ij})-\frac{3}{7}(\dot{\sigma}_{R}^{ijk}%
\dot{\sigma}_{R}^{ijk}+\dot{\sigma}_{I}^{ijk}\dot{\sigma}_{I}^{ijk}), \nonumber\\
\dot{P}^{i} &=&-\frac{2}{15}\dot{\sigma}_{R}^{kl}\dot{\sigma}%
_{I}^{jl}\epsilon ^{ijk}-\frac{\sqrt{2}}{7}(\dot{\sigma}_{R}^{jk}\dot{\sigma}%
_{R}^{ijk}+\dot{\sigma}_{I}^{jk}\dot{\sigma}_{I}^{ijk})\nonumber\\
&&-\frac{3}{7}\dot{\sigma}_{R}^{klm}\dot{\sigma}_{I}^{jlm}\epsilon ^{ijk}. \\
\dot{J}^{i} &=&\frac{1}{5}(\sigma _{R}^{kl}\dot{\sigma}_{R}^{jl}+\sigma
_{I}^{kl}\dot{\sigma}_{I}^{jl})\epsilon ^{ijk}\nonumber\\
&&+\frac{9}{7}(\sigma _{R}^{klm}\dot{\sigma}_{R}^{jlm}+\sigma _{I}^{klm}\dot{\sigma}_{I}^{jlm})\epsilon
^{ijk}.
\end{eqnarray}

Since the r.h.s of the above equations are quadratic in the radiation terms we only need a linear relationship between the radiation data and the PN multipole expansion. Using the linearized Einstein's equation in the TT gauge and following \cite{Blanchet}, one finds that
\begin{eqnarray*}
\sigma _{R}^{ij} &=& -\sqrt{2} U^{ij} \\
\sigma _{I}^{ij} &=& \frac{8}{3\sqrt{2}}V^{ij} \\
\sigma _{R}^{ijk} &=& -\frac{1}{9}U^{ijk} \\
\sigma _{I}^{ijk} &=& \frac{1}{6}V^{ijk}
\end{eqnarray*}

Thus, both have identical r.h.s. to this order. This is a remarkable result since the evolution equations come from completely different approaches. On the other hand one must be careful with the final equations of motion for the center of mass, energy and spin of the system since their relationship to kinematical variables are different in both formulations. In several PN papers, the recoil velocity of the center of mass is defined as $\frac{\Delta P^i}{M}$ which is the integral of eq. (\ref{ppunto}) divided by the total mass. However, it follows from eq. (\ref{momento}) that in our formulation one obtains a different result. This a straightforward consequence that in this formalism the gravitational radiation is part of the total linear momentum. In some sense this is analogous to the definition of momentum in electrodynamics where the kinematical definition $\Sigma_i m_i \overrightarrow{v}_i$ as well as the electromagnetic radiation enter in the definition of $P^i$. A more careful look into these differences will be addressed in the future.
\section{Final Comments and Conclusions}
We summarize our results and make some final remarks.
\begin{itemize}
\item We have defined the notion of center of mass and spin for asymptotically flat spacetimes, i.e., spacetimes where there is a precise notion of an isolated gravitational system.

\item The main tools used in our approach are the linkages together with a canonical NU foliation constructed from solutions to the Regularized Null Cone cut equation. The RNC cut foliation is given in the so called Newman Penrose gauge with a vanishing shear in the asymptotic past. Physically this corresponds to an isolated gravitational system which is asymptotically stationary in the past and it is specially useful to describe the emission of sources like of closed binary coalescence, supernovas or scattering of compact objects.

\item The RNC cut equation is an important ingredient in this construction. Its 4-dim solution space together with the lorentzian metric constructed from the solutions of the RNC cut equation provide the background to define the center of mass worldline. In this work we have used a perturbative approach to the RNC cut equation to introduce a flat metric at the zeroth order and a first order solution of the RNC cut equation to obtain NU foliations, one for each timelike worldline (with respect to the flat metric) on the solution space.

\item We have obtained the center of mass worldline by requiring that the mass dipole moment vanishes at the special NU foliation associated with this worldline. We have derived equations of motion for the center of mass and spin linking their time evolution to the emitted gravitational radiation. They are given by a very simple set of equations that resemble their Newtonian counterparts and thus should be useful in generalizing many well known results in astrophysics when very energetic processes are considered.

\item In astrophysics very often one assumes conservation laws for isolated systems. However, our equations show that for highly energetic processes were a fair percentage of energy is emitted as gravitational radiation, this is far from being true. We have shown here that this radiation affects the motion of the center of mass and the spin of the system and the solutions to the above equations yield their dynamical evolution.

 \item We have compared our approach with the ANK formulation to check for differences and similarities. This comparison suggest that our definitions of mass dipole moment and angular momentum are better suited to allow for higher multipole radiation or spacetimes with rotational symmetry.

 \item We have also compared our equations with those derived from the PN formalism. Although we have mainly done so for a very simple set of global variables, the results are very encouraging since, to second order approximation, the r.h.s. of the evolution equation for these variables are identical in both formulations. However, the relationship between total linear momentum and the velocity of the center of mass is different in both approaches. This difference might disappear after taking a careful look at other variables like radiative vs local shears, etc. A lot more work is needed to find a bridge between these formulations that start at opposite ends, one at null infinity, the other from local definitions based on the sources.

 \item It is believed that in late 2017 aLIGO will be operational to detect radiation from coalescing neutron stars and/or black holes. As we are all aware, numerical waveforms will never be able to fill out the parameter space needed for a coincidence check. Therefore, several ODE models like the PN approach or the EOB have been pursued with that goal in mind. Our approach should be useful to the ODE models for the reasons outlined below.

     The PN approach has several tentative definitions of center of mass with vanishing acceleration while emitting gravitational radiation. Since the motion of the center of mass is crucial in analyzing the motion of the coalescing sources, evolution of the mass and current momenta, and finally in the plot of the waveform in time domain, it is important to know whether or not the center of mass has an acceleration during this process. In this work we have shown that the center of mass has an acceleration which is partially given by the radiation reaction of $M$, $P^i$, and $J^i$ and partially given by the relationship between the center of mass velocity, gravitational radiation and the global quantities $M$, $P^i$, and $J^i$. Following our results, the equations of motion for the coalescing sources should be revised if it can be shown that the waveform changes when the center of mass has acceleration. This will be addressed in the future.
 \item Finally we want to address an important conceptual issue, the meaning of the observational space, i.e. the solution space of the RNC cut equation.

 In this work we have used worldlines on a 4-dim Minkowski space constructed on the solution space of the RNC cut equation. This flat metric can be regarded as the zeroth order approximation on a perturbation procedure on NSF to construct Lorentzian metrics. Note also that, both the in ANK and PN approaches, a flat background metric is used to introduce worldlines and propagate the gravitational radiation. Thus, in the three formulations that provide equations of motion for the gravitational sources one uses the same Minkowski background to compare results.

 However, the RNC cut equation provides a method to construct a 4-dim observational space with a regular metric constructed from fields at null infinity. In which sense is the solution space points $x^a$ and regular metric associated with the RNC cut equation related to the "real spacetime" from which the gravitational radiation is obtained at null infinity?

 If the gravitational source is composed of ordinary matter. Then the NSF equations provides in principle a method to construct null cone cuts for "real" points of the spacetime. The equation has three different terms, a Huygens part made of gravitational radiation, a gravitational tail and a source term that includes integrals along spacetime lines and is responsible for caustics and singularities. Therefore, if one is able to detect gravitational waves one can then safely assume that is not on a caustic region. Moreover, we do not have the technology to detect gravitational tails and we conclude that the dominant part of the NSF equation for the situation assumed above is the Huygens one.

The RNC cut equation is the smoothed version of the NSF equation, obtained by neglecting the other contributions and extending de validity of the Huygens part to the whole sphere. Thus, one could define a "norm" for metrics constructed from the NSF and RNC cut equations using energy methods to see how far apart are the solutions. One should mention that this comparison is a highly non trivial task that is worth addressing in the future.

 Even if the space time contains black holes our approach can also assign a center of mass worldline. From the gravitational radiation reaching null infinity one constructs the observation space with a regular metric and in that space one defines the center of mass worldline associated with this radiation. In this case the regular metric of the solution space has no relationship to the spacetime with black holes. Nevertheless, from the gravitational radiation reaching null infinity one computes the equation of motion including the back reaction effects. If a black hole is formed after the coalescence, one can also compute its final position and velocity although one knows that a black hole evolution is not a worldline in the real spacetime. We find this a desirable feature of this formalism since it gives a method of defining particle worldlines without the infinities that appear when one introduces delta functions in stress energy tensors. It has been pointed out that this second method yields ill defined quantities \cite{Geroch-Trashen}.
\end{itemize}

{\bf Acknowledgements:} We are grateful to two anonymous referees for their questions/comments that help to improve the quality of this paper. This research has been supported by grants from CONICET and the Agencia Nacional de Ciencia y Tecnolog\'ia.

\appendix
\section{Null Cone Cuts} \label{Appendix A}
We present here some properties of the NC cuts coming from a worldline on the spacetime.

  \begin{itemize}
  \item Globally the NC cuts are projections from smooth 2-dim Legendre submanifolds of the projective cotangent bundle of $\scri^+$ \cite{KLR,IKR,FNS}. It follows from this property that a generic NC cut has a finite number of singularities and those singularities can be classified as either cusps or swallowtails. Thus, locally the NC cuts are smooth 2-surfaces at null infinity.
 \item For the gravitational systems we would like to describe, compact sources such in the observational volume space of aLIGO, it is always possible to give a local description of the cuts in a given Bondi coordinate system as
    \begin{equation}
u_B=Z(x^a,\zeta, {\bar \zeta}).
\end{equation}
\item The above equation has also a second meaning, namely, for fixed values of $(u_B,\zeta, {\bar \zeta})$ the points $x^a$ that satisfy the above equation form the past null cone from the point $(u_B,\zeta, {\bar \zeta})$ at null infinity. Thus, $Z$ satisfies
    \begin{equation} \label{g00}
g^{ab}(x)\partial_a Z  \partial_b Z= 0,
\end{equation}
\item Under a Bondi supertranslation $\tilde{u}_B = u_B +\alpha (\zeta,{\bar \zeta})$, $Z$ transform as $\tilde{Z} = Z +\alpha$,. However, neither the conformal metric nor the field equations for $Z$ change under a supertranslation as they all depend on spacetime derivatives of $Z$.
\item The explicit algebraic construction of the conformal metric is done by first selecting a $((\zeta, {\bar \zeta})$ family of) null coordinate system $u=Z(x^a,\zeta, {\bar \zeta}), \omega = \eth Z, {\bar \omega}= {\bar \eth} Z, R =  {\bar \eth} \eth Z$ and then extracting the metric components from (\ref{g00} by successive $\eth$ and ${\bar \eth}$ derivatives of (\ref{g00}).

 \item It can be shown that all the non trivial components of the conformal metric are obtained in terms of spacetimes derivatives of a function $\Lambda(x^a,\zeta, {\bar \zeta})$ defined as
 \begin{equation}\label{Z}
\eth^2 Z= \Lambda(x^a,\zeta, {\bar \zeta}).
\end{equation}
This function $\Lambda$ plays a major role in the field equations for the NSF. Note that from its definition it follows that
\begin{equation}\label{Z}
{\bar \eth}^2 \Lambda= \eth^2 {\bar \Lambda}.
\end{equation}
This condition is called the reality condition and it will be used below to restrict the free data in the field equations.

\item A perfectly valid question is whether a conformal metric can be constructed from any arbitrary function $Z(x^a,\zeta, {\bar \zeta})$. In general the answer is no since for a fixed value of $x^a$ equation (\ref{g00}) is an algebraic equation for nine constants whereas  $(\zeta, {\bar \zeta})$ can take any value. Thus, conditions must be imposed on $Z$ for a metric to exist. It can be shown that the so called metricity conditions are given by
    \begin{equation} \label{m.2}
\eth^3 \left(g^{ab}(x)\partial_a Z  \partial_b Z\right)= 0,
\end{equation}
    and they must be satisfied by $Z$ before one looks for a conformal metric. These conditions generalize work by Cartan\cite{Cartan}, and Chern\cite{Chern} originally derived for third order ODEs although coming from a completely different approach. In three dimensions, one derives exactly the same condition from either the NSF\cite{FIK} or the Cartan approach.
\item Further insight into the geometrical meaning of $\Lambda$ can be gained by Using Sachs theorem. One can show that the $\Lambda$ satisfies
\begin{equation}\label{NC cuts}
\Lambda=\sigma^0(Z,\zeta,{\bar \zeta})-\sigma_Z(x^a, \zeta,{\bar \zeta}),
\end{equation}
with $\sigma^0$ the Bondi shear at null infinity and  $\sigma_Z$ the asymptotic shear of the future null cone from $x^a$ evaluated at null infinity \cite{BKR}.
\item Under a Bondi supertranslation the above equation remains valid in form as $\sigma'^0=\sigma^0+\eth^2 \alpha $ and $\sigma_Z$ remains the same.
\item As a particular case of equation (\ref{NC cuts}), one can obtain the null cone cuts  in Minkowski space. Since $\sigma_Z=0$ (null cones are shear free) and the imaginary part of the shear vanishes we have,
\begin{equation*}
\eth^2 Z_M=\sigma^0(\zeta,{\bar \zeta}) = \eth^2 \sigma_R ,
\end{equation*}
  with $\sigma_R$ a real function on the sphere. Using a supertranslation one eliminates $\sigma_R$ and obtains a canonical equation
  \begin{equation*}
\eth^2 Z_0=0 ,
\end{equation*}
whose solution will be given and used in this work.

   \end{itemize}

  To obtain the dynamical equations for $Z$ one uses the algebraic relation between the conformal metric of the spacetime and $Z$ directly from (\ref{g00}). One then constructs the Ricci and Weyl tensor and imposes the Einstein equations. Since the resulting equations are technically involved we present first the linearized version of the field equations for $\Lambda$ and $Z$.

  Keeping only terms of order $\Lambda$ in the metric components and writing down the Ricci flat equation to linear order in $\Lambda$ one gets
  \begin{equation*}
\Box \Lambda=0 ,
\end{equation*}
with $\Box$ the D'Alembertian in flat space. Thus, $\Lambda$ satisfies Huygens principle and its solution only depends on the data given on the flat null cone cut. If in addition one imposes the metricity conditions and the reality condition for $\Lambda$ one gets
\begin{equation}\label{linear cuts}
  \frac{\partial}{\partial u} \left[{\bar \eth}^2 \Lambda-{\bar \eth}^2 \sigma^0(u,\zeta,{\bar \zeta}) - \eth^2{\bar \sigma}^0(u,\zeta,{\bar \zeta})\right] =0,
\end{equation}
and as expected, the above equation is supertranslation invariant.

To go from the above equation to the RNC cut equation one replaces $\Lambda$ by $\eth^2Z$, $u$ by $Z$, etc. obtaining
\begin{equation}
\bar\eth^2\eth^2(Z-Z_i)=\bar\eth^2(\sigma-\sigma_i)+\eth^2 (\bar\sigma-\bar\sigma_i)
\end{equation}
with $Z_i$ some initial cut, and $\sigma_i=\sigma(Z_i,\zeta,{\bar \zeta})$. Thus, $Z-Z_i$ is supertranslation invariant and only depends on $\sigma-\sigma_i$.

Defining $[Z] = Z -Z_i$, $[\sigma^0]= \sigma^0 -\sigma^0_i$, one writes the formal linearized solution as

\begin{eqnarray}
[Z](x^a,\zeta,\bar{\zeta})= x^a \ell_a+ \oint K(\zeta,\zeta^{\prime})(\bar \eth^{\prime 2} [\sigma^{\prime 0}] + \eth^{\prime 2} [{\bar \sigma}^{\prime 0}])dS^{\prime 2},\nonumber\\
\end{eqnarray}
with $x^a \ell_a$ and $K(\zeta,\zeta^{\prime})$ the kernel and the Green function of the $\bar \eth^2 \eth^2$ operator on the sphere. In the above equation the four constants $x^a= (R^{0},R^{i})$ are interpreted as points in the spacetime whereas $\qquad \ell_a=(Y_0^0,-\frac{1}{2}Y^0_{1i})$ are the $l=0.1$ spherical harmonics. Also $\sigma^{\prime 0}=\sigma^0(x^a \ell^{\prime}_a,\zeta^{\prime},{\bar \zeta}^{\prime})$.

We now seek for a one parameter family of solutions that represents worldlines on the spacetime, i.e., $x^a= x^a(\tau)$. For this family we set $Z_i= Z(x^a(\tau_i),\zeta,{\bar \zeta})$. Instead of finding the most general form of the solution to the above equation we want to concentrate in the compact systems we are interested in describing. Essentially we would like to describe sources in the volume of space that can be observed by aLIGO such as closed binaries, supernovae or gravitational kicks. For those situations it is fair to assume that the system is asymptotically stationary both in the past and in the future. We thus assume that as $\tau_i \to -\infty$ the imaginary part of the Bondi vanishes and the cut $Z_i$ is shear free. In that limit we get $\sigma_i= \eth^2 \sigma_R(\zeta,{\bar \zeta})$, $Z_i= x^a(\tau_i)\ell_a+\sigma_R(\zeta,{\bar \zeta})$. Selecting a Bondi frame with vanishing $\sigma_R(\zeta,{\bar \zeta})$ we obtain
\begin{equation}\label{RNC cuts}
 {\bar \eth}^2 \eth^2 Z= {\bar \eth}^2 \sigma^0(Z,\zeta,{\bar \zeta}) + \eth^2{\bar \sigma}^0(Z,\zeta,{\bar \zeta}).
\end{equation}

This equation is used in this work and is referred to as the regularized null cone cut equation or RNC cut equation for short. Its linearized version was independently derived by L. Mason \cite{Mason} and by Fritelli and collaborators \cite{FKN2}.

Although the RNC cut does not corresponds to any spacetime point, the full NSF equation\cite{BKR} can be used to check how far apart is the RNC cut from a "real" cut coming from a spacetime point. Assuming the propagation is mostly along the characteristics and there are no caustics in the propagation (this is the type of situation aLIGO will be operating) then the main contribution to the NSF equation is the Huygens part. Work of Luc Blanchet show that the contribution of gravitational tails on binary coalescence are 5 to 7 orders of magnitude smaller than the leading part of the radiation. Thus, the real null cone cut is locally smooth and close to, in a precise way given by the non-Huygens terms of the NSF equation, a RNC cut.

This seems to be the case for the gravitational radiation that can be detected by aLIGO. If aLIGO can only detect radiation for null directions where  the intensity is higher it is safe to assume that for such isolated system the RNC cut will adequately describe the null cone from the center of mass since one only detects the Huygens part of the gravitational wave.

We thus claim that the solution space of the RNC cut equations is useful to describe the dynamical behaviour of global variables such as the center of mass and intrinsic angular momentum defined in this work.

\section{Octupole Contribution} \label{Appendix B}
To include the octupole contribution in our equation of $\dot{M}, \dot{P}^i$ and $\dot{J}^i$ we write the expansion of the Bondi shear of the set of eqs. (\ref{expan_tensorial}) in the form
\begin{equation}
\sigma^0=\sigma ^{ij}(u_{B})Y_{2ij}^{2}(\zeta,\bar \zeta )+\sigma ^{ijk}(u_{B})Y_{3ijk}^{2}(\zeta,\bar \zeta ).
\end{equation}
The energy and the linear momentum loss are the $l=0,1$ component of the eq. (\ref{psiprima}). Introducing the above equation in (\ref{psiprima}) and using the tensorial harmonics products table of refs. \cite{KQ} and \cite{ngilb} we get
\begin{eqnarray*}
\dot{M} &=&-\frac{c}{10G}(\dot{\sigma}_{R}^{ij}\dot{\sigma}_{R}^{ij}+%
\dot{\sigma}_{I}^{ij}\dot{\sigma}_{I}^{ij})-\frac{3c}{7G}(\dot{\sigma%
}_{R}^{ijk}\dot{\sigma}_{R}^{ijk}+\dot{\sigma}_{I}^{ijk}\dot{\sigma}%
_{I}^{ijk}), \\
\dot{P}^{i} &=&\frac{2c^{2}}{15G}\dot{\sigma}_{R}^{jl}\dot{\sigma}%
_{I}^{kl}\epsilon ^{ijk}-\frac{\sqrt{2}c^{2}}{7G}(\dot{\sigma}_{R}^{jk}\dot{\sigma}%
_{R}^{ijk}+\dot{\sigma}_{I}^{jk}\dot{\sigma}_{I}^{ijk})\nonumber\\
&&+\frac{3c^{2}}{7G}\dot{\sigma}_{R}^{jlm}\dot{\sigma}_{I}^{klm}\epsilon ^{ijk}.
\end{eqnarray*}

To obtain the quadrupole and octupole contribution to the angular momentum loss we compute the imaginary $l=1$ component of the definition (\ref{DJB}). For this we use (\ref{psi1prima}) and the tensorial harmonics products table to get
\begin{equation*}
\dot{J}^{i}=\frac{c^{3}}{5G}(\sigma _{R}^{kl}\dot{\sigma}_{R}^{jl}+\sigma
_{I}^{kl}\dot{\sigma}_{I}^{jl})\epsilon ^{ijk}+\frac{9c^{3}}{7G}(\sigma _{R}^{klm}\dot{\sigma}_{R}^{jlm}+\sigma _{I}^{klm}%
\dot{\sigma}_{I}^{jlm})\epsilon ^{ijk}
\end{equation*}
In a similar way one can write the n-pole contribution to the evolution equation for $M$, ${P}^{i}$ and ${J}^{i}$.

\section{The Tensor Spin-s Harmonics} \label{Appendix C}
In order to clarify the derivation of the main results in this article we give several tensor spin-s harmonics examples.
\begin{eqnarray*}
Y_{0}^{0} &=&1 \\
Y_{1i}^{0} &=&\eth Y_{1i}^{-1}=\bar{\eth }Y_{1i}^{1} \\
\eth Y_{1i}^{0} &=&-2Y_{1i}^{1} \\
\bar{\eth }Y_{1i}^{0} &=&-2Y_{1i}^{-1} \\
\eth Y_{1i}^{1} &=&0 \\
\bar{\eth }Y_{1i}^{-1} &=&0 \\
\eth \bar{\eth }Y_{1i}^{1} &=&-2Y_{1i}^{1} \\
\bar{\eth }\eth Y_{1i}^{-1} &=&-2Y_{1i}^{-1}
\end{eqnarray*}
and for $l=2$
\begin{eqnarray*}
Y_{2ij}^{1} &=&\bar{\eth }Y_{2i}^{2} \\
\eth Y_{2ij}^{1} &=&\eth \bar{\eth }Y_{2ij}^{2}=-4Y_{2ij}^{2} \\
Y_{2ij}^{0} &=&\bar{\eth }^{2}Y_{2ij}^{2} \\
\eth Y_{2ij}^{0} &=&-6Y_{2ij}^{1} \\
\eth \bar{\eth }Y_{2ij}^{0} &=&-6Y_{2ij}^{0}
\end{eqnarray*}
For more details and definitions of the Spin-weighted spherical harmonics the reader must see ref. \cite{ngilb} and for higher products see ref \cite{KQ}.

\end{document}